\pdfoutput=1
\documentclass[aps,prl,reprint,groupedaddress]{revtex4-1}
\usepackage[pdftex]{graphicx}
\usepackage{natbib}
\usepackage{amssymb,amsfonts,latexsym,color,dcolumn,bm,subfigure}

\usepackage{comment}

\begin{document}


\title{Near-field studies of anisotropic variations and temperature induced structural changes in a supported single lipid bilayer}


\author{Merrell A. Johnson$^{\dag,\S}$ and Ricardo S. Decca$^\dag$}
\affiliation{$^\dag$Department of Physics, Indiana University Purdue University Indianapolis, 402 North Blackford Street, Building LD154,
Indianapolis, Indiana 46202, USA \\ 
$^\S$Department of Physics Indiana University Purdue University Fort Wayne, 2101 E. Coliseum Blvd., Building KT126B, 
Fort Wayne, IN 46805, USA}


\date{\today}

\begin{abstract}

Temperature controlled Polarization Modulation Near-Field Scanning Optical Microscopy (PM-NSOM) measurements of a single supported $L_{\beta^{\prime}}$ 1,2-dipalmitoyl-sn-glycero-3-phosphocholine (DPPC) lipid bilayer are presented.  The  effective retardance ($\Delta S = \frac{2 \pi (n_e-n_o)t}{\lambda}$), where $t$ is the thickness of the bilayer and $\lambda$ is the wavelength of light used) and the direction of the projection of the acyl chains ($\theta $) were measured simultaneously. From $\Delta S$ the birefringence ($n_e-n_o$) of the bilayer was determined.   A change of $\Delta S$ $\sim$ (3.8 $\pm$ 0.3)~mrad at the transition temperature $T_m \sim$ 41$^{\circ}$C between the gel $L_{\beta^{\prime}}$ to liquid disorder $L_{\alpha}$ was observed in a single planar bilayer. This agrees well with previous values of ($n_e-n_o$) in the $L_{\beta^{\prime}}$ phase and translates to an assumed $\langle \phi \rangle $ $\sim$ 32$^{\circ}$ when T $<T_m$ and 0$^{\circ}$ when T $>T_m$. Evidence of supper heating and supper cooling are presented, along with a discussion of the behavior that occurs around $T_m$.
  
\end{abstract}


\maketitle

\section{Introduction}
Charting the complex interface of biological membranes is essential in determining how localized interactions contribute to overall cellular function.  Containing an elaborate mesh of membrane proteins, assorted lipids and domains of many types, regional lipid orientation within membranes can give us great insight on their interactions with neighboring molecules.  Before approaching the complexities of a true cellular membrane less sophisticated model membranes have been characterized using techniques ranging from NMR to various x-ray methods \cite{phtr8,Sampcharct6,Dppcret2}.  These traditional methods accurately predict structural properties of lipid membrane multi-stacks and give great insight in the physical behavior of these systems.  One drawback is that they do not yield information on low curvature/single membrane systems naturally found in biological cellular membranes.  Those techniques also are limited in the lateral information they are able to obtain across the lipid bilayer.  To address these limitations a high resolution technique that has an elevated degree of sensitivity to structural changes throughout planar membrane systems was implemented.  

In addition, many have chosen methods that mainly focus on exploiting the anisotropic nature of lipids to obtain structural information of the lipid molecules oriented in the membrane \cite{ Dppcret6, Dppcret8, Dppcret7, Dppcret4,Dppcret2}.  In particular, various studies have been used to measure the effective retardance of the membrane systems on 1,2-dipalmitoyl-sn-glycero-3-phosphocholine (DPPC) bilayers in the $L_{\beta^{\prime}}$ gel state \cite{intro10, Dppcret4, blip8}.  Under these conditions an average constant molecular tilt, $\phi\, \sim \,32^\circ$ \cite{Dppcret5}, across the structure can be assumed. Knowing the perpendicular component of the refractive index $n_{\bot}$ of the acyl chains, the optical orientation $\theta $ and parallel refractive index $n_{\|}$ of the acyl chains can be determined.  All mentioned techniques were used in studying both bulk and planar lipid membrane systems, but give limited lateral information.  To address this issue Lee \textit{et al.} \cite{intro10} were able to explore these structural effects by measuring the effective retardance by using a Near-Field Scanning Optical Microscope (NSOM).  Doing so they were able to obtain a lateral resolution on the order of $\sim$ 100 nm across a planar membrane system.  While obtaining a high lateral sensitivity with NSOM, the measurement was limited in its sensitivity of the retardance,  leaving room for improvement on the detectability of the acyl chain tilt variations across planar membrane systems.    

In studying lipid systems it is also common to investigate their thermodynamic properties to better understand the localized interactions that take place between molecules.  Including temperature controlled NMR and x-ray experiments \cite{NMR1,NMR2,Saxs1}, techniques like differential scanning calorimetry  \cite{PTran2} have been used to study the first order phase transitions in lipid systems.  In this paper an attempt to address these limitations by introducing a polarization modulating technique \cite{NSOMref7} in combination with a temperature controlled Near-Field Scanning Optical Microscope is introduced.  Doing so a high degree of sensitivity of the acyl chain tilt, with a lateral resolution $\sim$ 100 nm has been obtained.  As a consequence of the enhanced a more accurate difference in index of refraction on the  membrane's plane is readily available.  The combined use of these techniques with temperature control allow to study the structural variations involved in first order phase transitions  planar bilayer systems. 
\vfill

\section{Experimental Details}

\subsection{Sample Preparation}
 
 \begin{figure}[b]
  \centering

  \subfigure[] {\includegraphics[width=.9\linewidth,height=.6\linewidth]{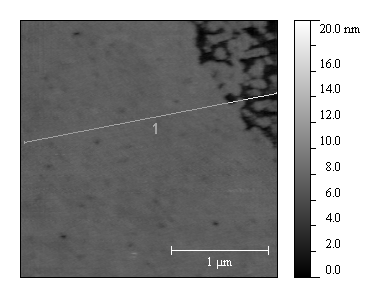}}
  \hspace{12pt}
  \subfigure[] {\includegraphics[width=.9\linewidth,height=.6\linewidth,trim=40 10 40 40, clip = true]{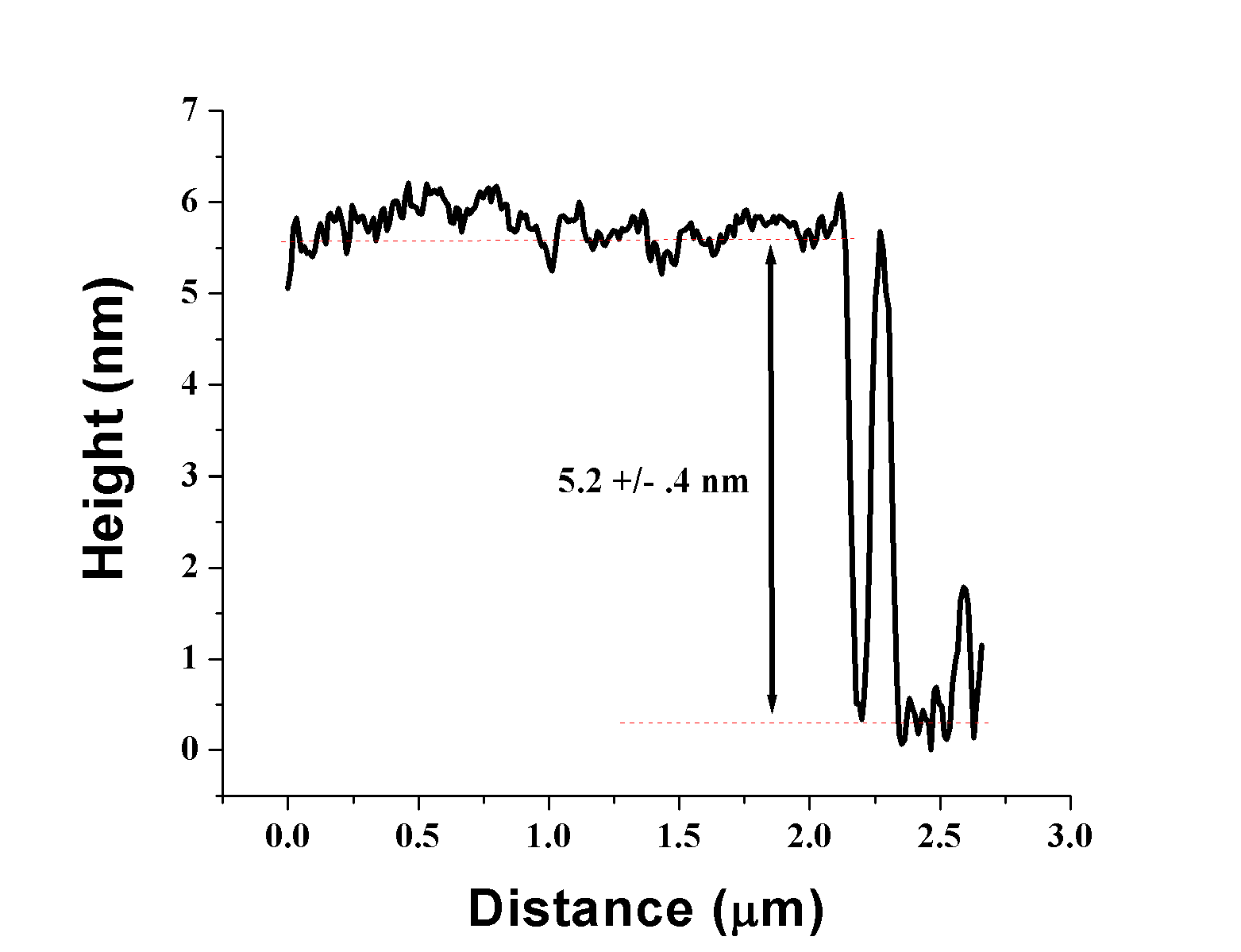}}\\
  \caption{AFM image of supported DPPC bilayers near an edge. (a) 2.6 $\mu$m $\times$ 2.6 $\mu$m topographical contrast image. The low height regions represent the underlying glass slide  (b) Plot of the height data taken from a line cut of the image in a), showing a height difference of $\approx$ 5.2 nm. }
  \label{afmchar}
  \end{figure}

Hydrophilic glass substrates were prepared by sonicating Fisher brand glass cover slips in detergent and deionized (DI) water separately for $\sim$ 15$-$20 minutes each.  This was followed with a wet-chemical oxidation process using a piranha solution (H$_2$SO$_4$~:~H$_2$O$_2$ = 3:1) for $\approx$ 5 min \cite{Sampprep2, Sampprep4}, to create a flat hydrophobic surface.  The cover slips were then rinsed and sonicated for $\sim$ 30 min in DI water and finally thermally dried under normal atmospheric conditions.
	
Supported lipid bilayers were formed on the prepared hydrophilic glass substrates by vesicle fusion \cite{Sampprep1, Sampprep4}.  1.0 - 0.5 mg/ml of 1,2-dipalmitoyl-sn-glycero-3-phosphocholine (DPPC) in $100 \text{ mM NaCl}~:~30 \text{ mM NaH}_2\text{PO}_4$ was sonicated at $\approx 60^{\circ} \text{C}$ until clear to create  small unilamellar vesicles (SUV’s).  100 $\mu$L of solution was placed on the glass substrates and left to equilibrate at room temperature for $\approx $ 30 mins.  The samples were then baked at $60^{\circ} \text{C}$ for 45 to 60 min.  After rinsing the samples with DI water, they were either re-hydrated and characterized with an AFM under a fluid cell (see figure \ref{afmchar}), or kept in a chamber at 100\% relative humidity until imaged with the PM-NSOM system.

\subsection{Experimental Setup and Supporting Details}

\begin{figure}[!]
\begin{center}
\subfigure[]{\label{MNSOMset1}}{\includegraphics[width=.95\linewidth,height=.7\linewidth]{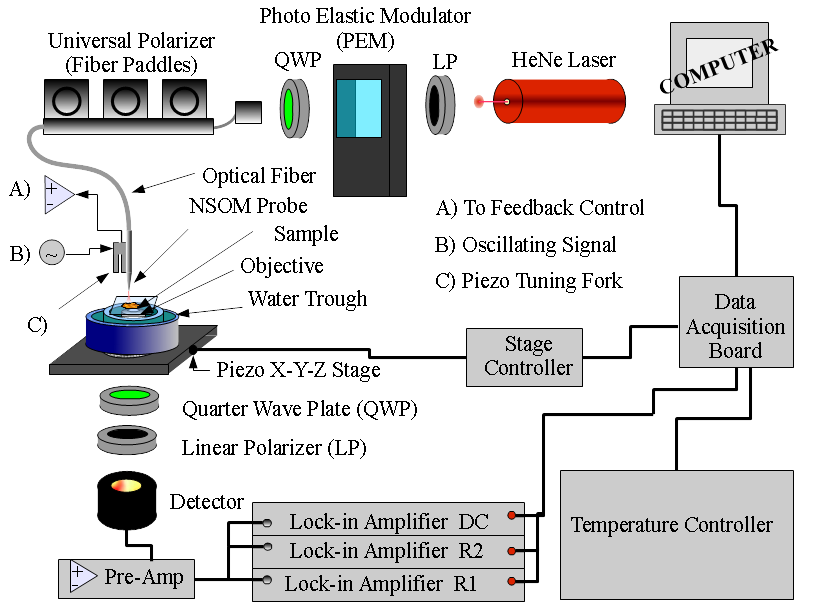}}\\
\subfigure[]{\label{MNSOMset2}}{\includegraphics[width=.95\linewidth,height=.8\linewidth,trim = 0 0 0 0, clip]{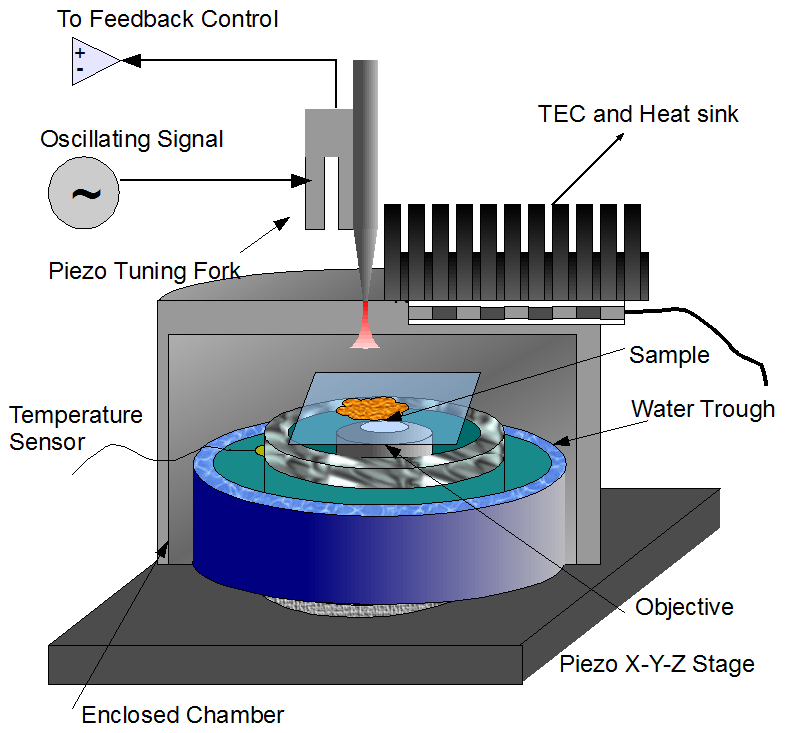}}
\subfigure[]{\label{MNSOMset3}}{\includegraphics[width=.49\linewidth,height=.6\linewidth,trim = 0 0 0 10, clip]{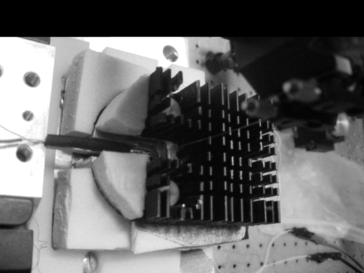}}
\subfigure[]{\label{MNSOMset4}}{\includegraphics[width=.49\linewidth,height=.6\linewidth,trim = 0 0 0 40,clip]{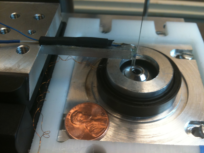}}
\caption{Diagram of PM-NSOM setup. (a) PM-NSOM setup highlighted with components for room temperature studies (b) Diagram of the temperature controlled chamber with water trough to maintain 100 \% relative humidity. (c,d) Photographs of the temperature controlled chamber.}
\label{PMNSOMset}
\end{center}
\end{figure}

	The total PM-NSOM setup is schematically shown in figure \ref{PMNSOMset}. All measurements were conducted with NSOM probes coated with $\sim$ 5 nm of Cr and $\approx$ 150-200 nm of Al, with diameters ranging from $\approx$ 80 to 100 nm (verified with a scanning electron microscope).  Monochromatic light from the (5-6 mW) helium neon laser ($\lambda \,=$~632.8~nm) was mechanically chopped and passed through a linear polarizer. All other optical elements are positioned with respect to this direction defined to be 0$^{\circ}$.  After the linear polarizer, light passed through a photo elastic modulator PEM (45$^{\circ}$, frequency 41.9 kHz) and QWP (0$^{\circ}$ ).  The light was then coupled into a single mode optical fiber, which was connected to an universal fiber polarizer, used to control the polarization going into the NSOM probe \cite{NSOMref7}.  The NSOM   tip-to-sample distance ($\sim$ 10 nm) was controlled utilizing a shear force feedback system \cite{NSOMref14,NSOMref15}. The light from the NSOM probe and sample was collected with an objective lens, and sent through a QWP (0$^{\circ}$), an analyzer (-45$^{\circ}$) and finally collected by a Si PIN diode detector.  The current from the diode was converted to a voltage signal by means of a transimpedance amplifier and measured  by three lock-in amplifiers (LIA). The three LIA's were locked at the frequency of the mechanical chopper (proportional to the intensity of the laser, DC value), the first and second harmonics of the PEMs frequency respectively. 

In the absence of a sample, it was initially assumed that the universal polarizer negates any polarizing effects from the fiber and NSOM probe. Under these circumstances, a Jones matrix formalism was applied and it was determined that the signal at the detector is proportional to $ \frac{1}{2}-\frac{1}{2}\sin (A)\sin (S) \cos (2 \theta )  +\frac{1}{2} \cos (A) \sin (S)\sin (2 \theta )$, where $A = B \cos \left(\omega t\right)$, $B$ is the magnitude of the retardance and $\omega $ is the angular frequency set by the PEM (The PEM was calibrated such that the $J_{0}(B)=0$ ).  The lock-in amplifiers are used to obtain a signal normalized by the DC term at the first harmonic ($I_{\omega}$) and the second harmonic ($I_{2\omega}$), 
\begin{equation}
I_{\omega} = \gamma J_1(B) \sin (S) \cos (2 \theta)
\label{intens1}
\end{equation} 
\begin{equation}
I_{2\omega} = \gamma J_2(B)  \sin (S) \sin (2 \theta) 
\label{intens2}
\end{equation}
\noindent Hence, 
\begin{equation}
\theta  =\frac{1}{2} \arctan \left(\frac{- I_{2\omega}\text{  }J_1(B)}{I_{\omega} J_2(B)}\right)
\end{equation}
\noindent and the retaradance $S$ is,
\begin{eqnarray}
S &=& \arcsin \left(\frac{ I_{\omega}}{ \gamma J_1(B)\cos (2 \theta )}\right)\nonumber  \\
&=&\arcsin \left(\frac{ -  I_{2\omega}}{ \gamma J_2(B)\sin (2 \theta )}\right) 
\end{eqnarray}
\noindent where $\gamma$ is a the term associated with the lock-in measurement (i.e. $\gamma =\,\frac{\pi}{2}$ due to the DC component being a square wave for our study).  During the alignment process in the absence of a sample, both $I_{\omega}$ and $I_{2\omega}$ in equations \ref{intens1} and \ref{intens2} are minimized, which implies that the retardances associated with the NSOM probe and fiber approach zero.

Moving the sample in and out of the optical path while aligning the NSOM system is not ideal due to the user's risk of damaging the probe and the cumbersome process of attaching another.  It would be advantageous to have a method where the alignment could be made over the sample of unknown retardance.   To understand this effect, the sample and probe were modeled as two independent retarding elements, with retardances $S$ and $\delta$ inserted between the two QWP in the previously described setup.  Following the same Jones matrix formalism  the expected intensities of the two retarding objects $S$ and $\delta$ oriented at different angles were calculated.  During the alignment process, where $I_{\omega}$ and $I_{2\omega}$ are minimized, the we can assume the orientation of the NSOM probe/fiber and sample are made equivalent.  Since all retardances are small, the intensities,
    \begin{equation}
I_{\omega} = \gamma J_1(B) \sin (S+\delta) \cos (2 \theta)
\end{equation}
and
\begin{equation}
I_{2\omega} = \gamma J_2(B)  \sin (S+\delta) \sin (2 \theta) 
\end{equation}

\noindent are obtained. 
From these expression, it is seen that instead of an absolute value of $S$, one can align the system over the sample and obtain a deviation in the retardance $\Delta S$ and a relative orientation $\theta$ across the sample.  

To ensure the correctness of our calibration approach and the viability of our previous calculations, PM-NSOM measurements were conducted on cleaved stepped muscovite (mica) substrates.  The mica substrate showed steps of different heights corresponding to different number of layers of the crystal.  Both far-field and near-field measurements were made of the retardance $S$ on mica, the far field measurements used to make sure NSOM determinations were accurate.  Both aforementioned approaches for measuring the retardance and orientation using the NSOM apparatus were performed.  It was observed that on average the overall difference  was $\Delta S \sim$ 0.3 mrad between the two methods. Using the height topography from the NSOM measurement  the birefringence of the muscovite crystal was determined to be $\sim$ 0.0025, which falls within accepted values \cite{mica1, mica2}.  It was noted that the images taken with the system aligned over the sample, yielded more detail in $\Delta S$ than just measuring $S$.  We suspect that this increase in sensitivity to sample variations is a consequence of the  measurements being done from a minimum instead of an offset signal, highlighting an additional advantage of aligning the system over the sample.

All experiments were conducted at 100$\%$ relative humidity, to ensure membrane structure contained the characteristic acyl chain tilt of $\approx$ 32$^{\circ}$ in the $L_{\beta^{\prime}}$ phase at room temperature.  The measurements were taken with scan times ranging from 100 ms to 450 ms per point, where $I_{1\omega}$, $I_{2\omega}$, $\Delta S$, $\theta $ and topography data were collected using a data acquisition board and LabView computer program. Trace and retrace images were collected to confirm measurement reliability. 

Temperature controlled PM-NSOM measurements were performed similarly to those conducted at room temperature but including the following procedures.  A cylindrical aluminum chamber with a small slit in one side to accommodate for the NSOM probe holder was placed around the NSOM probe (figure \ref{PMNSOMset}) to minimize the lateral heat gradients across the sample.  The chamber was thermally isolated with an insulating foam casing and the temperature was controlled using a Peltier thermoelectric cooler (TEC) plate to move heat in and out of the chamber.  The temperature of the sample was measured using a sensor  embedded into the aluminum sample holder. The sample holder rested on a Delrin stand with a stainless steel bottom to improve mechanical stability.  With this design the temperature stability was within $\pm \textbf{ } 0.07 ^{ \circ}$C.  A water trough surrounded the sample holder, which preserved the required 100$\%$ relative humidity.  

During a typical run  the chamber's temperature was first heated beyond $T_m$ to ensure the lipid membrane was well into the $L_{\alpha}$ state.  The NSOM probe was then engaged with the sample.  The chamber was cooled below $T_m$ and heated back to the initial temperature.  This was proven to be the most reliable method to counter the probe crashing events that occurred due to the thermal expansion of the Delrin separator.  

\section{Results}

The anisotropic nature of lipids yields differences in the index of refraction (or birefringence ($n_e - n_o$)) along orthogonal directions in the membrane's plane.  Because the polarizability of the lipid molecule is asymmetric the refractive indices along the length of the acyl chains ($n_{\parallel}$) and perpendicular ($n_{\perp}$) are not equal.  In this study, the principal optical axis is assumed to lie parallel to the length of the acyl chains $\phi \sim$ 32$^{\circ}$ with respect to the membranes' normal for DPPC.  Polarized light of wavelength $\lambda$ propagating in the $z$-direction, parallel to the membranes' normal, experiences a retardance between the $x$ and $y$ components of the electric field according to \cite{intro10} 
\begin{eqnarray}
\Delta S &=& \frac{2\pi (n_e - n_o)t}{\lambda} \nonumber \\
&=& \frac{2\pi t}{\lambda} \frac{1}{\sin \left( \arctan \left(\frac{n_{\|}{}^2}{n_{\bot }{}^2}\cot (\phi )\right)+\phi \right)} \nonumber \\ && \cdot \frac{n_{\bot }n_{\|}}  {\sqrt{n_{\bot }{}^2\sin^2 (\phi )+n_{\|}{}^2\cos^2 (\phi )}}-n_{\bot }.
\label{biref2}
\end{eqnarray}
\noindent The extraordinary ray corresponds to the electric field parallel to the optical axis, where as for the ordinary ray the electric field is perpendicular to it. The two rays were indistinguishable for the collection optics used, due to the overall thickness of the membrane.

\begin{figure}[!]
\begin{center}

\subfigure{\label{DPPCMT2S1}\includegraphics[width=.95\linewidth,height=.7\linewidth, trim = 0 0 0 0, clip]{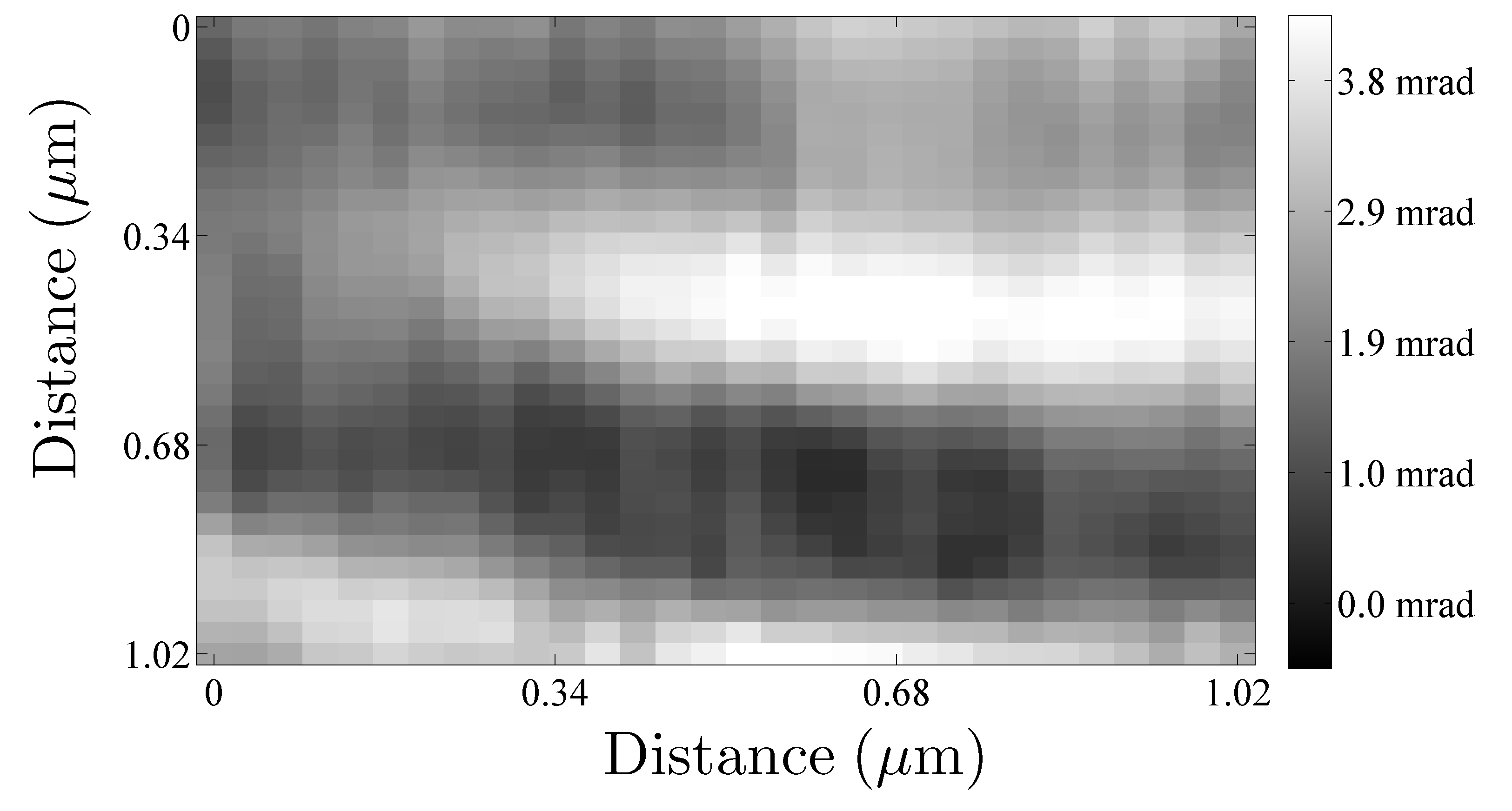}}
\hspace{24 pt}
\hspace{24 pt}
\subfigure[]{\label{DPPCMT2S4}\includegraphics[width=.95\linewidth,height=.6\linewidth,clip=true,trim= 270 0 10 160,clip]{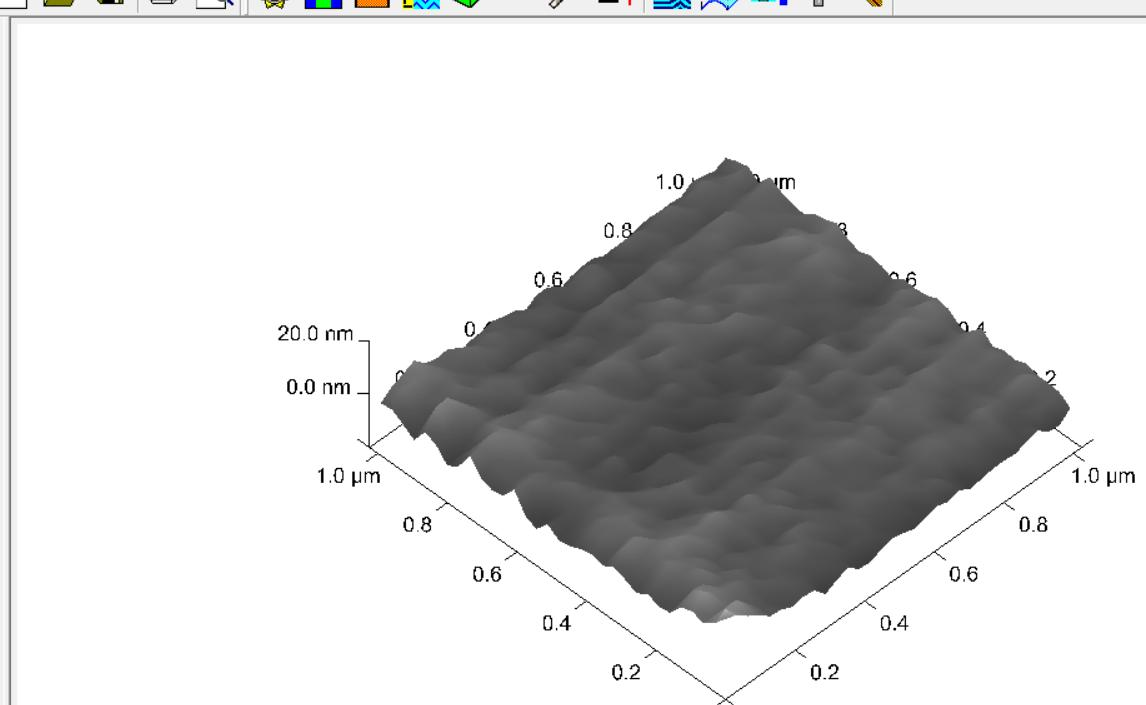}}
\caption{PM-NSOM picture of DPPC supported on glass, (a) $S$ as a function of position. (b) Topography obtained over the same area.}
\label{Dppcretard1}
\end{center}
\end{figure}

Topographic measurements were made across the sample until a discontinuity in $S$ and $\theta$ were found.  From the image of $S$ in figure \ref{DPPCMT2S1}, a measured $\Delta S $ of $\approx$ 3.9 $\pm$ 0.4 mrad was obtained by taking the average $S$ inside the hole and subtracting from a region away from the determined edge.  A birefringence $(n_e \,-\, n_o)\,=\, \frac{\Delta S \lambda}{2\pi \textit{t}} \approx$ 0.073 $\pm$ 0.008, using $\lambda$ = 632.8 nm and \textit{t} $=$ 5.2 nm $\pm$ 0.4 \cite{Dppcret3} was obtained.  These results agree well with previous measurements with a significant reduction in error \cite{intro10}. 
Furthermore an increase in $\Delta S$ to $\approx\,$ 0.0075 mrad was obtained at the edge of the hole. It is expected that $\phi$ is greater where the membrane forms a boundary.  Since the size of the hole is smaller than the probe's diameter, it did not show a significant change in height, see figure \ref{DPPCMT2S4}.

\begin{figure}[!]

\begin{center}
\subfigure{\label{DPCTmpS}\includegraphics[width=.95\linewidth,height=.7\linewidth, trim = 0 10 0 0, clip]{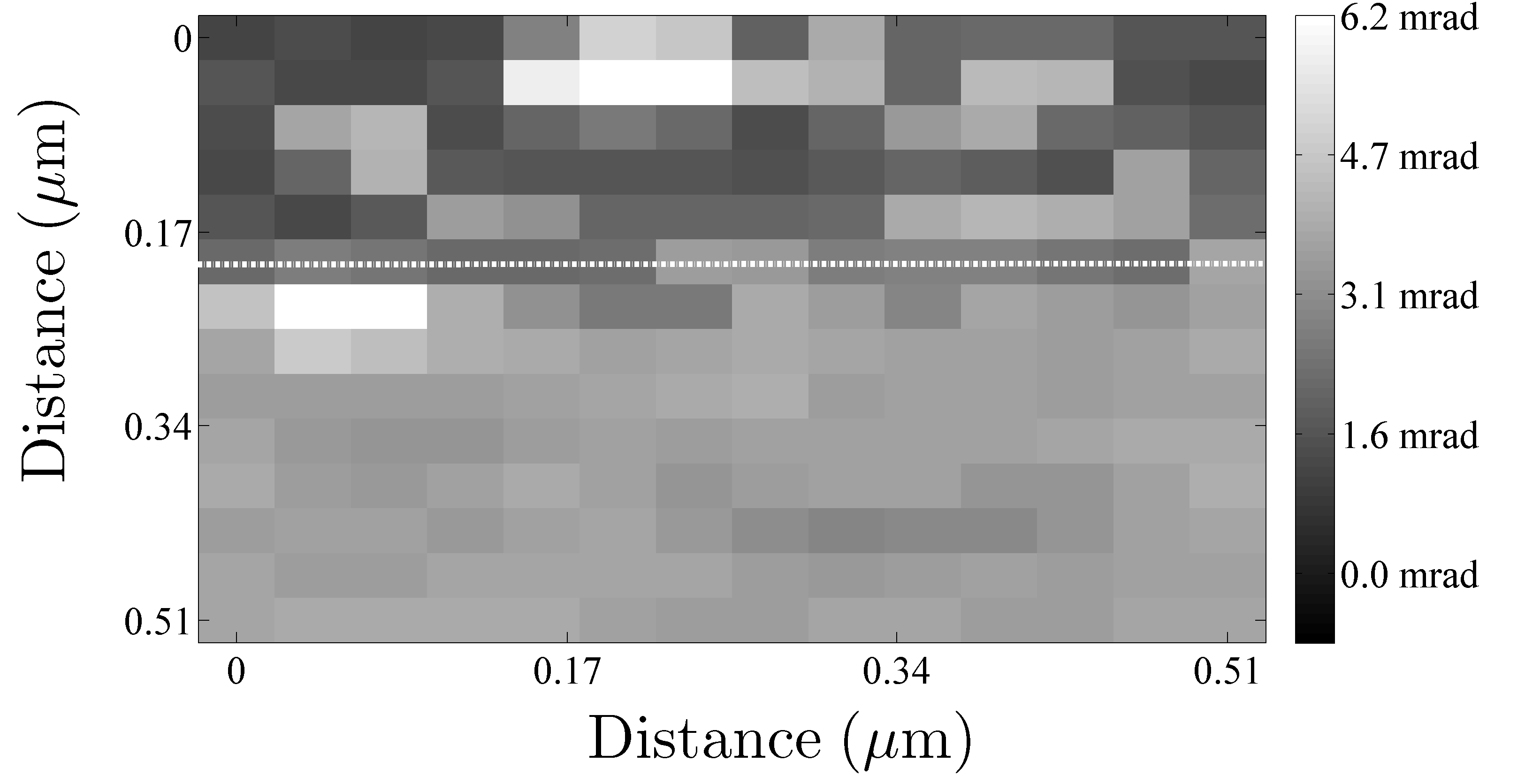}}\\
\hspace{36pt}
\caption{Temperature controlled PM-NSOM 512 nm $\times $ 512 nm images of DPPC supported on glass. An image of the measured $S$, showing a $\Delta $ $S$ of $\approx $ 3.8 $\pm$ 0.3 mrad at the $T_m$, which is highlighted by the white dotted line.  This change $S$ indicates a change of $\phi $ of the acyl chains.}
\label{temp1}
\end{center}
\end{figure}

\begin{figure}[!]
\begin{center}
\subfigure[]{\label{DPRetGrallH}\includegraphics[width=.95\linewidth,height=.7\linewidth,trim= 20 0 90 50,clip]{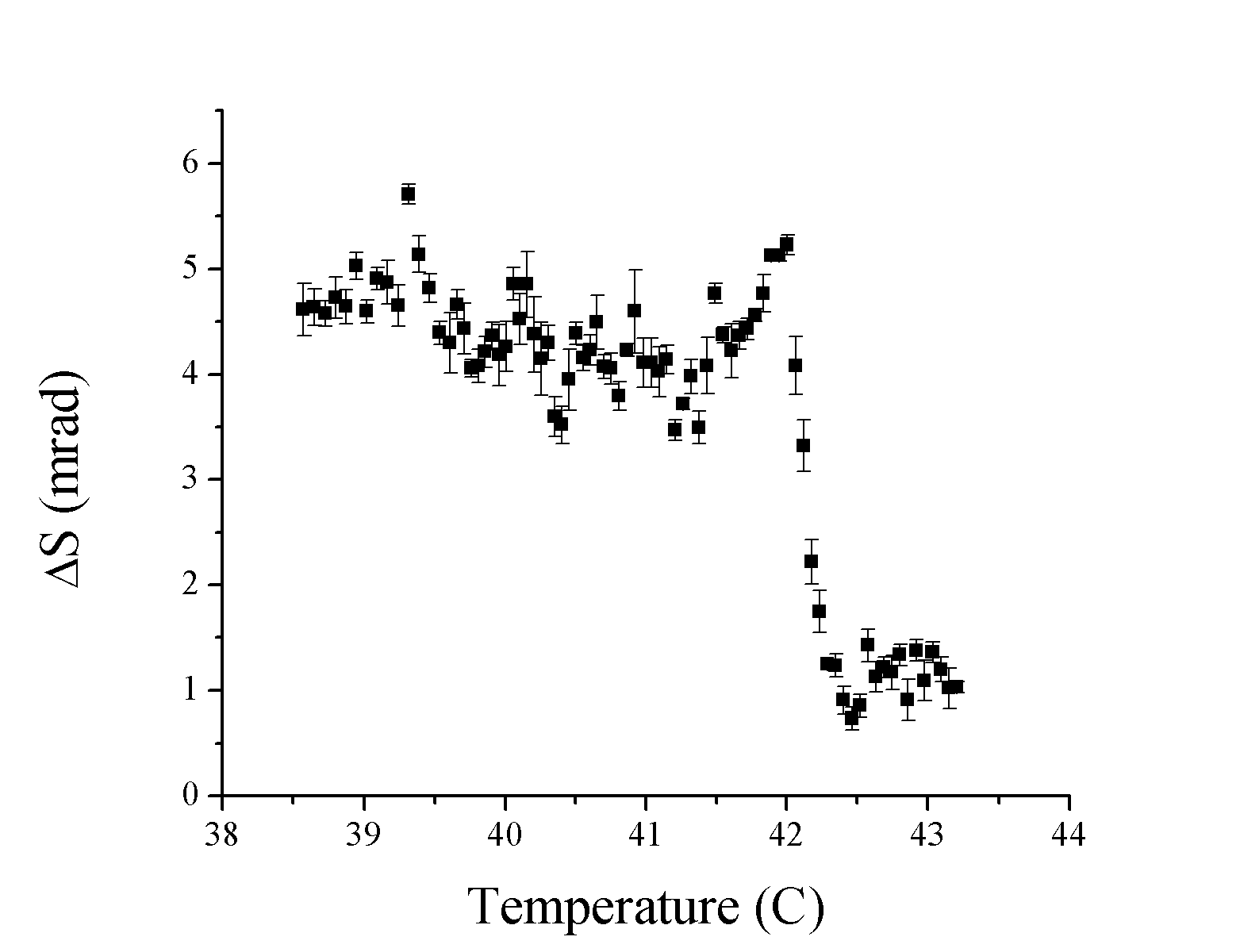}}\\
\subfigure[]{\label{DPRetGrH}\includegraphics[width=.95\linewidth,height=.7\linewidth, angle = 0,trim= 0 0 90 50,clip]{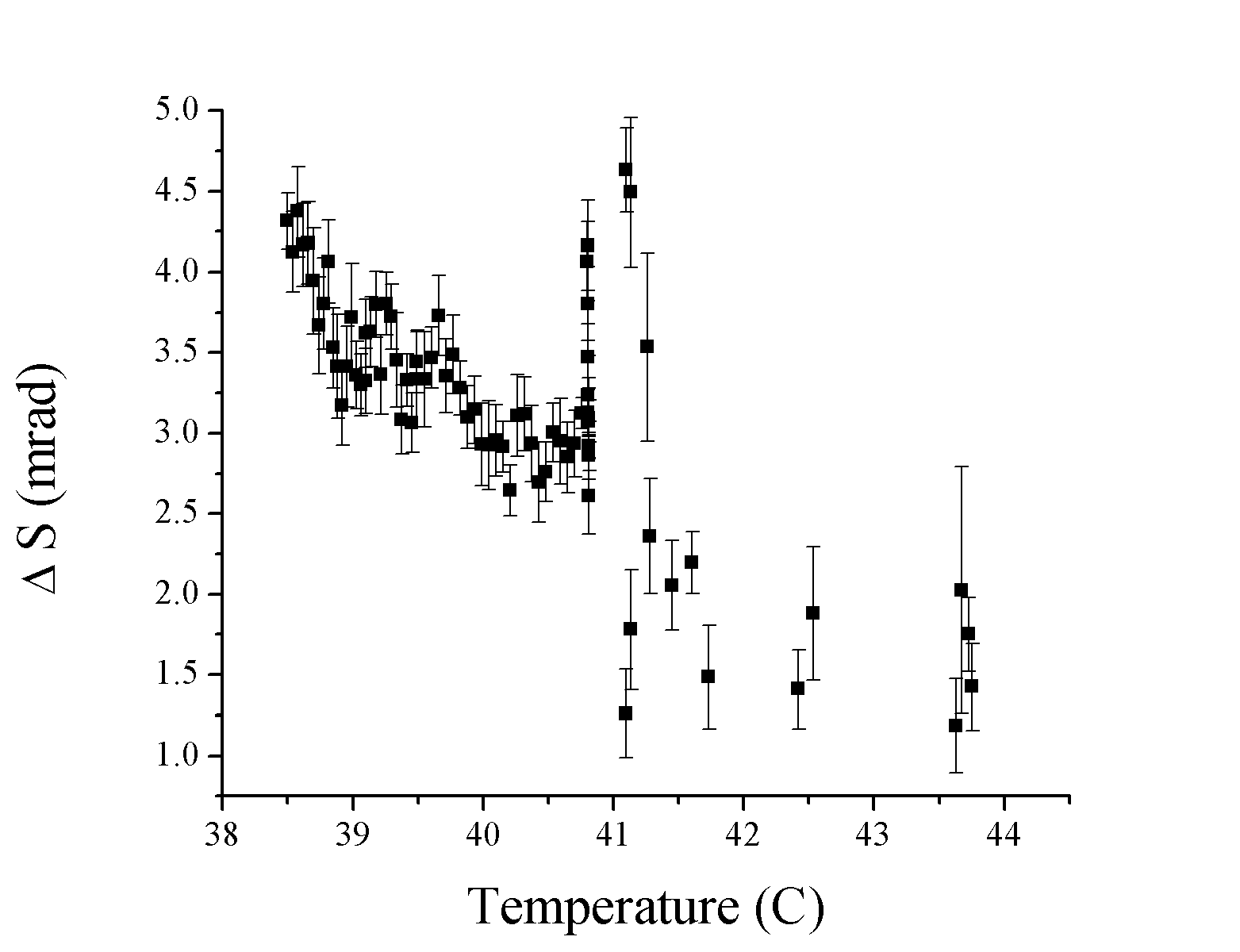}}\\
\caption{$ \Delta S$ values from a temperature controlled PM-NSOM measurement of DPPC supported on glass. (a) Graph of $\Delta S_{avg} $ versus temperature as the system was heated, taken from (512 nm)$^2$ images showing a change in $S$ across $T_m$ of $\approx $ 4.0 $\pm$ 0.4 mrad.  (b) Cooling data graph of $\Delta S_{avg} $ versus temperature, taken from $\sim$ (100 nm)$^2$ images.}
\label{DppcRetversustmp}
\end{center}
\end{figure}

A series of temperature PM-NSOM experiments probed the main phase transition temperature $T_m \sim 41^{\circ}$C of DPPC across a 512 nm $\times$ 512 nm area.  The 16 $\times$ 16 pixel images were taken over $\approx$ 3.5 min at a rate of $\sim$ 0.06 $^{\circ}$C$/$min.  The series started at $T \approx 38^\circ$ and finishing at $T \approx 42^\circ$.  Figure \ref{DPCTmpS}, highlights the main transition occurring over on image time scale.  In the figure, $S$ is observed to remain constant at low $T$ and then it shows a jump of $\Delta S \approx  \text{(}3.84  \pm 0.20 \text{) } \text{mrad}$ at $\approx$ 41.1$^{\circ}$C. This $\Delta S$ is interpreted as the average position of the acyl chains change from their characteristic $\langle \phi \rangle$ of $\approx 32^{\circ}$ in the $L_{\beta^\prime}$ to the $\langle \phi \rangle$ $\rightarrow$ 0 in the $L_{\alpha}$ state.  A change of $\approx 21^{\circ}$ at $T_m$ was observed in $\theta$ at the same position highlighted in figure \ref{DppcRetversustmp}.  This change in  $\theta$ corresponds to the sample in the $L_{\beta^\prime }$ state with optical system having one $\theta$, and changing to a different value, characteristic of the optical system, when $T > T_m$.  

The next series of temperature controlled experiments decreased the image acquisition time to $\approx $ 50 s, and the temperature change per image was $\sim$ 0.05$^{\circ}$C at a rate $\sim$ 3 $^{\circ}$C/min.  This was done to explore the hysteresis effects of the 1$^{st}$ order phase transition of the planar membrane system.  Images taken were identical in size to the previous described temperature controlled experiments.  The results for $S$ are seen in figure \ref{DppcRetversustmp}, where a $\Delta S$ of $\approx$ 3.5 mrad is observed across $T_m$.  Figure \ref{DPRetGrallH} Shows that $S \approx 4.7$~mrad for  $T < T_m$ and $\approx $ 1 mrad when $T > T_m$.  This translates to the acyl chains' $\langle \phi \rangle$ transitioning from $\sim$ 32$^{\circ}$ to zero, calculated using equation \ref{biref2}.  This is what we expect to  occur as the lipid bilayer goes from the $L_{\beta^\prime }$ into the L$_\alpha$ phase.

\section{Discussion}

	With the knowledge obtained from the PM-NSOM experiments at room temperature on supported DPPC in the $L_{\beta^\prime }$ phase,  physical parameters about the lipid molecules in the  the membrane can be extracted.  Birefringence yields  information on the polarizability of the lipid molecules.  Following the model of rigid cylinders used by Salamon \textit{et al}. \cite{Dppcret4}  equation \ref{biref2} and the relationship $ n_i{}^2= \frac{\alpha _i\epsilon _o}{\left(V -\alpha _iL_i\right)}+ \epsilon _o,$ (where $V$ is volume and $L_i$ is the shape factor for a cylinder \cite{AnisoMat1}) can be used to determine the transverse ($\alpha_{t}$) and longitudinal ($\alpha_{l}$) polarizabilites.  The values for two acyl chains was determined to be, $\alpha_{t} \, = \, 44.2 \text{\AA}^3$ and $\alpha_{l}\, =\, 94.4 \text{\AA}^3$, assuming the area per lipid ($A$) and acyl chain length ($l$) to be 47.9 $\text{\AA}^2$ and $17.2\,\text{\AA}$ respectively \cite{Dppcret5}.  These values are very close to the theoretically calculated polarizabilities $\alpha_{t}$ = 25.14 ${\text{\AA}}^3$ and $\alpha_{l}$ = 45.8 $\text{\AA}^3$ \cite{Dppcret4, AnisoMat1} of a single palmitic acid C$_{16}$ .

\begin{figure}[!]
\begin{center}
\subfigure{{\includegraphics[width=.95\linewidth,height=.7\linewidth, angle = 0]{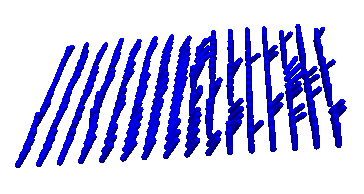}}}\\
\subfigure{{\includegraphics[width=.95\linewidth,height=.5\linewidth, scale = .1,trim= 120 0 120 0,clip]{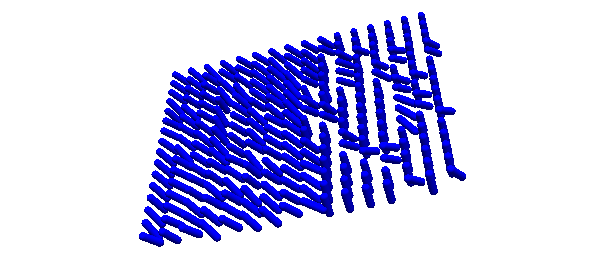}}}
\caption{ 3D representation of the average acyl chain orientation throughout the membrane. The imag was reconstructed from the data in figure \ref{DPCTmpS})  Each tube portrays the average position of many lipid molecules contained in a $\sim$ (100 nm)$^2$ area.}
\label{DPPC512chainmap}
\end{center}
\end{figure}

	Beyond extracting physical parameters, the model previously presented can be used  to create a three dimensional representation of the average direction of the orientation of the acyl chains of the lipid molecules within the membrane.  Using the values for $n_\bot$ and $n_\|$, the average orientaion $\phi$ of the acyl chains across the NSOM probe's aperture with respect to the membrane's normal.  Figure \ref{DPPC512chainmap} utilized the data taken from figure \ref{temp1}, representing the molecules average orientation as blue rods.  As shown, when the membrane is below $T_m$, $\langle \phi \rangle $ was $\approx $ 32$^{\circ}$ while when it is heated above the phase transition temperature it was in the $L_{\alpha}$ phase, where $\langle \phi \rangle $ is zero.  
	
\begin{figure}[!]
\centering
\subfigure[]{\label{fluct1}}\includegraphics[width=.95\linewidth,height=.7\linewidth, angle = 0,trim= 0 30 20 10,clip]{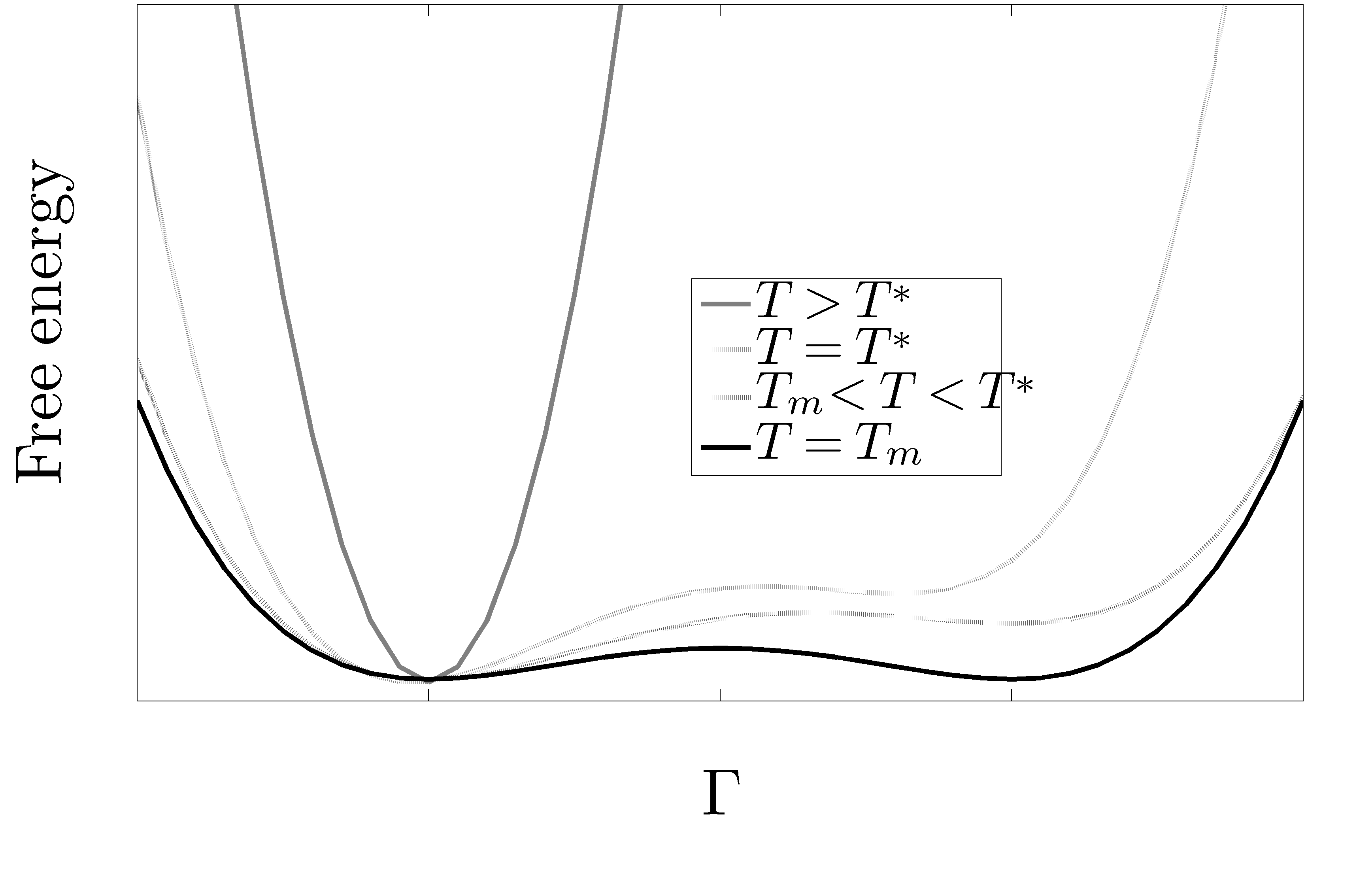}
\subfigure[]{\label{fluct2}}\includegraphics[width=.95\linewidth,height=.7\linewidth, angle = 0,trim= 0 10 90 20,clip]{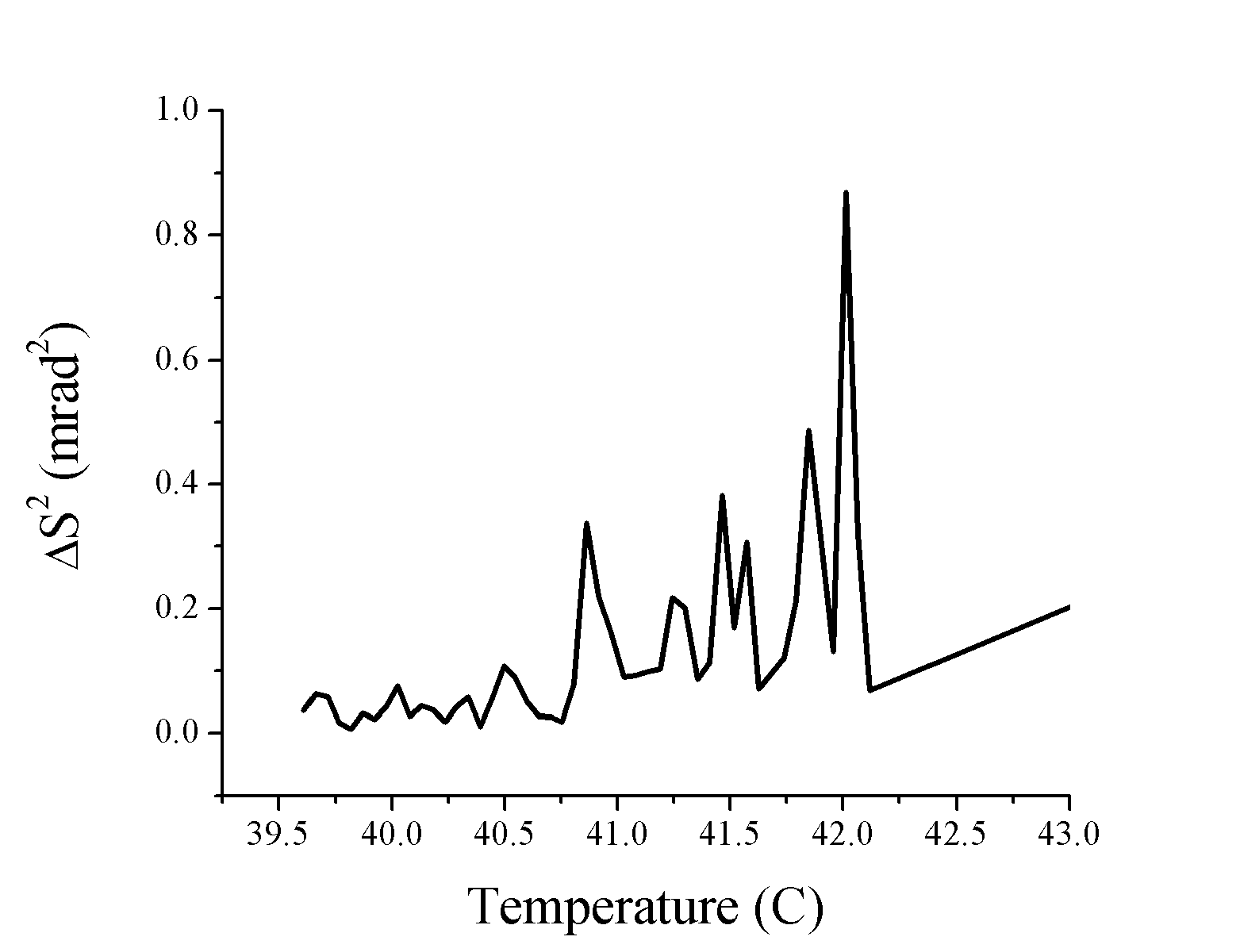}

\caption{(a) Graphical depiction of the free energy of a system as a function of the order parameter($\Gamma$). (b) Graph of $\langle \Delta S^2 \rangle$ versus $T$, extracted by analyzing the various of the image in the heating data presented in figure \ref{DPRetGrallH}.}  
\label{retvar1}
\end{figure}

 	  Through further analysis from the data taken in figure \ref{DppcRetversustmp}, we observed that $\langle S \rangle $ varies as $T_m$ was approached.  This observation could derive from varying orientation that exist in the $P_{{\beta}^\prime }$ phase.  A characteristic that cannot be confirmed with PM-NSOM.  Previous experiments on on multi-layered bilayer systems have measured the "ripples" and their periodicity, which are $<<$ than the diameter of the NSOM aperture \cite{blip9,blip10}.
 	  The hysteresis observed under the applied heating rates is comparable to what has previously been reported \cite{PTran1, PTran2}.  What was not initially evident  was that as $T_m$ was approached, the overall variability in $S$ increased, displayed in the graph of the variance ($\Delta S^2$) versus $T$ in figure \ref{fluct2}.  Within a Landau-Ginzburg picture of the first-order phase transition in nematic-to-isotropic systems \cite{phtr12} the observations can be understood.  It is expected that as the system crosses the melting temperature ($T_m$) and approach the supper-cooled or super-heated temperature ($T^*$) the shape of the free energy curve changes, as shown in figure \ref{fluct1}, where the free energy is depicted as a function of the order parameter $\Gamma$. In the process of heating, as $T$ increases past $T_m$, the shape of the free energy functional and the width of the metastable minimum slightly increases, allowing for an increase in the fluctuations of $\Gamma$ and consequently of any thermodynamic quantity.  This increase of deviations from the mean in $\Gamma$ is then correlated to the subtle increase in $\langle \Delta S^2\rangle$ as the system approached $T^*$.  The increase in the variance in $S$ correlates with an increase in state variability from figure \ref{fluct1}.  It can be clearly seen that before the transition $\langle \Delta S^2\rangle$ is smaller in magnitude than when the system approaches $T^*$ ($\sim 42 ^{\circ}$C).  This may suggest that this observation highlights these fluctuations in the metastable region predicted in the Landau-Ginzburg model, which will be investigated further in future work.

\section{Conclusions}
PM-NSOM was utilized in determining the anisotropic structural properties in supported lipid bilayers.  The technique was shown to improve previous measurements on $S$ by increasing the sensitivity and independently determining the direction of the projection of the acyl chains onto the membrane's surface.  With that increased sensitivity,  the longitudinal and transverse polarizability can be more accurately determined yielding results comparable to their theoretically calculated values.  It was also shown that one can obtain accurate lateral high resolution $\Delta S$ and orientation images with the PM-NSOM system while aligning the system over the sample.  This proved to be imperative in conducting the temperature controlled experiments and allowed one to measure relative values of $S$ and $\theta$.  

By adding temperature control to PM-NSOM the main phase transition from the gel state to the liquid disorder phase was observed. The melting temperature of a single DPPC lipid bilayer was found to be in good agreement with previously reported work.  The information from $S$ and $\theta$ allowed to create a three dimensional model of the average orientation of the lipid molecules within the membrane.  The system's sensitivity and control allowed for the observations of increased fluctuations  as  the metastable region of the phase diagram of the lipid membrane was probed.  

Future work could extend into membrane systems containing lipid mixtures or even protein/lipid complexes, where the properties of domains with different phases or orientations may be exploited.  With temperature control, the lateral dynamics of phase separations or even changes in the orientation of the lipid bilayer through regions of interest could be observed.  Overall it was shown how PM-NSOM can be utilized as a highly sensitive/non-invasive technique to study single lipid bilayer systems and achieve structural information beyond the ability of conventional optical techniques.

\begin{acknowledgments}
\section{acknowledgments}
We would like to acknowledge the use of the atomic force microscope associated with the Integrated Nanosystems Development Institute (INDI) and Nanoscale Imaging Center (NIC) at Indiana University Purdue University Indianapolis.  

\end{acknowledgments}

\bibliographystyle{apsrev4-1}
\bibliography{TDPPCFinal}

\end{document}